\documentclass[pra,10pt,showpacs]{revtex4}
\usepackage{graphicx,amsmath,amssymb}

\newcommand{\beq}{\begin{equation}}
\newcommand{\eeq}{\end{equation}}
\newcommand{\beqa}{\begin{eqnarray}}
\newcommand{\eeqa}{\end{eqnarray}}

\begin{document}

\title{Quantum Violation: Beyond Clauser-Horne-Shimony-Holt Inequality}

\author{Hoshang Heydari}
\email{hoshang@imit.kth.se} \affiliation{Institute of Quantum
Science, Nihon University, 1-8 Kanda-Surugadai, Chiyoda-ku, Tokyo
101-8308, Japan}

\date{\today}

\begin{abstract}
The best upper bound for the violation of the
Clauser-Horne-Shimony-Holt (CHSH) inequality was first derived by
Tsirelson.  For increasing number of $\pm 1$ valued observables on
both sites of the correlation experiment, Tsirelson obtained the
Grothendieck's constant ($\mathcal{K}_{G}\approx 1.73\pm0.06$) as a
limit for the  maximal violation. In this paper, we  construct a
generalization of the CHSH inequality with four $\pm 1$ valued
observables on both sites of a correlation experiment and show that
the quantum violation approaching $1.58$. Moreover, we estimate the
maximal quantum violation of a correlation experiment for large and
equal number of $\pm 1$ valued observables on both sites. In this
case, the maximal quantum violation converges to
$\sqrt{3}\approx1.73$ for very large $n$, which coincides with the
approximate value of Grothendieck's constant.
\end{abstract}


\maketitle

\section{Introduction}
The seminal paper of Einstein, Podolsky, and Rosens (EPR)\cite{EPR},
and  Schr\"odinger's article \cite{schroed} on quantum correlations
of entangled states as well as Bell \cite{Bell64} subsequent
discovery that quantum theory is incompatible with any locally
realistic, hidden variable theory have generated substantial
discussions and many experiments on the nature of quantum
non-locality. The violation of Bell's inequality was the first
mathematically sharp criterion for entanglement. A quantum state is
said to be unentangled, separable if and only if it can be written
as a convex combination of product states. In some cases, however,
this criterion fails to detect any entanglement. The standard
example of the Bell inequality is the CHSH inequality \cite{CHSH},
which refers to correlation experiments with two $\pm 1$ valued
observables on two sites. In this paper, we will only discuss the
CHSH type inequality. However, there is an infinite hierarchy of
such Bell type inequalities, which can basically be classified by
specifying the type of correlation experiments they deal with. The
CHSH inequalities are by far the best-studied cases of Bell
inequalities. The essential assumption leading to any Bell
inequality is the existence of a local realistic model, which
describes the outcomes of a certain class of correlation
measurements. Various aspects of the hierarchy of Bell inequalities
have already been investigated. Garg and Mermin\cite{Garg82}, for
instance, have resumed the idea of Bell and discussed systems with
maximal correlation. Gisin \cite{Gisin99} investigated setups with
more than two dichotomic observables per site with arbitrary states,
which we will also discuss in the following section. $N$-particle
generalizations of the CHSH inequality were first proposed by
Mermin\cite{Mermin90}, and further developed by
Ardehali\cite{Ardehali92}, Belinskii and Klyshko\cite{Klyshko93},
and others\cite{Roy91,Gisin98}. The best upper bound for the
violation of the CHSH inequality, first derived by
Tsirelson\cite{Tsirelsonbound}, is obtained by squaring the Bell
operator and utilizing the variance inequality\cite{Landau}. In the
case of more than two dichotomic observables per site only very
little is known about the limit. In particular there is yet no
explicit characterization of the extremal inequalities, although
constructing some inequalities, e.g. by chaining CHSH
inequalities\cite{BraunsteinCaves} is not difficult. However,
Tsirelson\cite{Tsirelsonbound,Tsirelsonbound1} recognized that the
quantum correlation functions, which are in general rather
cumbersome objects, can be reexpressed in terms of finite
dimensional vectors in Euclidean space.  For two observables on one
site and an arbitrary number on the other, Tsirelson showed that the
maximal quantum violation is $\sqrt{2}$. However, for an increasing
number of observables on both sites, he obtained the upper bound for
Grothendieck's
 constant $\mathcal{K}_G$ ($\approx 1.78$), known from the
geometry of Banach spaces, as the limit for the maximal
violation\cite{werner}. In particular,
 $\mathcal{K}_G$ is the smallest number, such that,
for all integers $n\geq 2$, all $n\times n$ real matrices
$[a_{ij}]$, and all $s_1,\cdots,s_n$, $t_1,\cdots,t_n\in\mathbf{R}$
such that $|s_i|,|t_j|\leq 1$  for which
$|\sum_{i,j}a_{ij}s_it_j|\leq1$, it is true that
$\Big|\sum_{i,j}a_{ij}\langle x_i,y_j\rangle\Big|\leq
\mathcal{K}_G$, where $x_1,\cdots,x_n$, $y_1,\cdots,y_n$ such that
$\|x_i\|,\|y_j\|\leq 1$ are vectors in a real Hilbert space.
Tsirelson \cite{Tsirelsonbound} showed that comparisons between
probabilities in classical physics and probabilities in quantum
mechanics yield discrepancy measures $\mathcal{K}_n$ for finite
$n\times n$ real matrices that approach Grothendieck's constant
$K_G$ for very lager $n$. The exact value of $\mathcal{K}_G$ is
unknown. A lower bound of $\pi/2$ was established by Grothendieck
\cite{Grothendieck}. In a recent paper, P. C. Fishburn and J. A.
Reeds \cite{Fishburn} showed that $\mathcal{K}_{q(q-1)}\geq
(3q-1)/(2q-1)$ for $q\geq 2$ and $\mathcal{K}_{20}\geq
\frac{10}{7}$; $n=20$ is the smallest known $n$ for which
$\mathcal{K}_n>\sqrt{2}$.
In this paper, we will construct a generalization of the CHSH
inequality with four observables on both sites  and show that
maximal quantum violation approaches $\sqrt{\frac{5}{2}}\simeq1.58$.
Moreover, we will estimate the maximal quantum violation for very
large numbers of observables per site in a correlation experiment.

\section{The structure of the set of quantum correlations}
In this section, we will define CHSH inequality and Tsirelson
inequality as the best upper bound for the violation of the CHSH
inequality. First, let us define CHSH inequality as follows. Let
$\mathrm{Cor}_{\mathcal{C}}(n,m)$ denote the set of classically
representable matrices, whose matrix elements are
\begin{equation}
\langle\mathrm{X}_{k},\mathrm{Y}_{l}\rangle_{c}=
\int\mathrm{X}_{k}(\varphi) \mathrm{Y}_{l}(\varphi) d(\varphi),
\end{equation}
where $\mathrm{X}_{k},\mathrm{Y}_{l}$ are random variables
satisfying $|\mathrm{X}_{k}|\leq1,|\mathrm{Y}_{l}|\leq1$. Then, the
CHSH inequality is defined by
\begin{equation}\label{chsh1}
|\langle\mathrm{X}_{1},\mathrm{Y}_{1}\rangle_{c}+
\langle\mathrm{X}_{1},\mathrm{Y}_{2}\rangle_{c}
+\langle\mathrm{X}_{2},\mathrm{Y}_{1}\rangle_{c}
-\langle\mathrm{X}_{2},\mathrm{Y}_{2}\rangle_{c}|\leq2.
\end{equation}
The CHSH inequality holds for any local-realistic theory. However,
quantum correlation violate the CHSH inequality, that is, let us
consider the following observables
\begin{equation}
\widehat{\mathrm{X}}_{k}=\widehat{\mathrm{X}}^{(1)}_{k}\otimes\mathrm{I}^{(2)}
~\text{and}~
\widehat{\mathrm{Y}}_{l}=\mathrm{I}^{(1)}\otimes\widehat{\mathrm{Y}}^{(2)}_{l}
\end{equation}
for all $k=1,2,...,n$ and $l=1,2,...,m$, where $\mathrm{I}^{(j)}$ is
identity operator on the Hilbert space $\mathcal{H}_{j}$, such that
the following relations are satisfied by these operators $
[\widehat{\mathrm{X}}_{k},\widehat{\mathrm{Y}}_{l}]=0$ for all $k$
and $l$ that is $\widehat{\mathrm{X}}_{k}$ is compatible with each
$\widehat{\mathrm{Y}}_{l}$. Hence for an arbitrary state
$\rho\in\mathcal{H}_{1}\otimes\mathcal{H}_{2}$, the quantum
correlation is measurable and
$\|\widehat{\mathrm{X}}_{j}\|\leq1,\|\widehat{\mathrm{Y}}_{k}\|\leq1$
for all $k$ and $l$. Thus we can define the quantum correlation
matrix $\mathcal{C}$ as
\begin{equation}
\mathcal{C}=
(\langle\widehat{\mathrm{X}}_{k},\widehat{\mathrm{Y}}_{l}
\rangle_{\rho})_{k=1,2,...,n;~l=1,2,...,m},
\end{equation}
where
$\langle\widehat{\mathrm{X}}_{k},\widehat{\mathrm{Y}}_{l}\rangle_{\rho}=\mathrm{Tr}(\rho
\widehat{\mathrm{X}}_{k}\widehat{\mathrm{Y}}_{l})$. Now, let the
convex set $\mathrm{Cor}_{\mathcal{Q}}(n,m)$ be quantum
-representable matrices of some quantum observables
$\widehat{\mathrm{X}}_{k},\widehat{\mathrm{Y}}_{l}$ as describe
above. The geometrical description of this convex set follows from
the following theorem \cite{Holevo,Tsirelson}: The matrix
$\mathcal{C}$ belongs to the set $\mathrm{Cor}_{\mathcal{Q}}(n,m)$
if and only if there exist vectors
$\mathrm{a}_{1},\mathrm{a}_{2},...,\mathrm{a}_{n}$ and
$\mathrm{b}_{1},\mathrm{b}_{2},...,\mathrm{b}_{m}$in Euclidean space
of dimension $\min(n,m)$, such that $\|\mathrm{a}_{k}\|\leq1$,
$\|\mathrm{b}_{l}\|\leq1$ and $ \mathrm{a}_{k}\cdot\mathrm{b}_{l}=
\langle\widehat{\mathrm{X}}_{k},\widehat{\mathrm{Y}}_{l}\rangle_{\rho},
$ for all $k$ and $l$.

Now, let us define for $n=m=2$ the Bell operator with  the same
structure as the combination  which appears on the CHSH inequality
as
$\mathcal{B}_{2,2}=\widehat{\mathrm{X}}_{1}\widehat{\mathrm{Y}}_{1}
+\widehat{\mathrm{X}}_{1}\widehat{\mathrm{Y}}_{2}
+\widehat{\mathrm{X}}_{2}\widehat{\mathrm{Y}}_{1}
-\widehat{\mathrm{X}}_{2}\widehat{\mathrm{Y}}_{2}$. Then, we have
\begin{equation}
\mathcal{B}^{2}_{2,2}=4
\mathrm{I}-[\widehat{\mathrm{X}}_{1},\widehat{\mathrm{X}}_{2}]
[\widehat{\mathrm{Y}}_{1},\widehat{\mathrm{Y}}_{2}],
\end{equation}
where we have assumed
$\widehat{\mathrm{X}}^{2}_{k}=\widehat{\mathrm{Y}}^{2}_{l}=\mathrm{I}$
for all $k$ and $l$. From this inequality we can get the CHSH
inequality as
\begin{eqnarray}\nonumber
\mathrm{Tr}(\rho
\mathcal{B}_{2,2})&=&\langle\widehat{\mathrm{X}}_{1},\widehat{\mathrm{Y}}_{1}\rangle_{\rho}+
\langle\widehat{\mathrm{X}}_{1},\widehat{\mathrm{Y}}_{2}\rangle_{\rho}\\&&+
\langle\widehat{\mathrm{X}}_{2},\widehat{\mathrm{Y}}_{1}\rangle_{\rho}-
\langle\widehat{\mathrm{X}}_{2},\widehat{\mathrm{Y}}_{2}\rangle_{\rho}\leq2,
\end{eqnarray}
whenever
$[\widehat{\mathrm{X}}_{1},\widehat{\mathrm{X}}_{2}]=
[\widehat{\mathrm{Y}}_{1},\widehat{\mathrm{Y}}_{2}]=0$.
The upper bound which gives the maximal violation of CHSH is
called Tsirelson inequality and is given by
\begin{equation}
\mathrm{Tr}(\rho \mathcal{B}_{2,2})\leq2\sqrt{2},
\end{equation}
where for any observable
$\widehat{\mathrm{Z}}=\widehat{\mathrm{X}},\widehat{\mathrm{Y}}$
and $\|\widehat{\mathrm{Z}}_{k}\|\leq
1,\|\widehat{\mathrm{Z}}_{l}\|\leq 1$ for all $k,l=1,2$, we have
estimate $|[\widehat{\mathrm{Z}}_{1},\widehat{\mathrm{Z}}_{2}]\|$
as follows
\begin{eqnarray}
|[\widehat{\mathrm{Z}}_{1},\widehat{\mathrm{Z}}_{2}]\|&\leq&
\|\widehat{\mathrm{Z}}_{1}\widehat{\mathrm{Z}}_{2}\|
+\|\widehat{\mathrm{Z}}_{2}\widehat{\mathrm{Z}}_{1}\|\\\nonumber&\leq&
\|\widehat{\mathrm{Z}}_{1}\|\|\widehat{\mathrm{Z}}_{2}\|
+\|\widehat{\mathrm{Z}}_{2}\|\|\widehat{\mathrm{Z}}_{1}\|\leq 2.
\end{eqnarray}
We will use this estimation in the next section when we derive an
inequality for the case $n=m=4$ and will try to estimate the maximal
violation of generalized CHSH inequality.
\section{Maximal quantum violation}
The maximal violation for an increasing number of observables on
both sites of a correlation experiment is still an unsolve problem.
However, Tsirelson has obtained the Grothendieck's constant as a
limit for the maximal violation. In this case we have
\begin{equation}
\mathrm{Cor}_{\mathcal{C}}(n,m)\subset\mathrm{Cor}_{\mathcal{Q}}(n,m).
\end{equation}
For example, the CHSH inequality provides a hyperplane separating
the polyhedron  $\mathrm{Cor}_{\mathcal{C}}(2,2)$ from the quantum
realizable matrix $\mathcal{R}_{2,2}=\left(%
\begin{array}{cc}
  1 & 1\\
  1 & -1 \\
\end{array}%
\right)$, such that
$\mathcal{R}_{2,2}\in\mathrm{Cor}_{\mathcal{Q}}(2,2)$. So it is
natural to ask how much $\mathrm{Cor}_{\mathcal{Q}}(n,m)$ exceeds
$\mathrm{Cor}_{\mathcal{C}}(n,m)$. Let $\mathcal{K}(n,m)$ be the
smallest number having this property, that is
\begin{equation}
\mathrm{Cor}_{\mathcal{Q}}(n,m)\subset\mathcal{K}(n,m)\mathrm{Cor}_{\mathcal{C}}(n,m).
\end{equation}
Then this sequence increases with $n$ and $m$. It was found by
Tsirelson, from geometrical description of the set
$\mathrm{Cor}_{\mathcal{Q}}(n,m)$, that
\begin{equation}
\mathcal{K}=\lim_{n,m\rightarrow\infty}\mathcal{K}(n,m)
\end{equation}
coincides with the Grothendieck's constant $
\mathcal{K}_{G}\leq{\pi\over2\ln(1+\surd2)}\approx1.78$ known from
the geometry of Banach spaces. In the next section, we will
construct an generalized CHSH inequality with more than two
observables per site and show that for an arbitrary state this
inequality has an upper bound which is lager than the upper bound
for CHSH inequality and it approaches the approximative
Grothendieck's constant.

\section{Beyond Tsirelson inequality}

In the case of correlation experiments with more than two dichotomic
observables per site only very little is known. So we will here go
beyond this limit by allowing four dichotomic observables per site.
In this case $(n=m=4)$, we will consider an inequality that provides
a hyperplane separating the polyhedron
$\mathrm{Cor}_{\mathcal{C}}(4,4)$ from the quantum realizable matrix
\begin{equation}
\mathcal{R}_{4,4}=\mathcal{R}_{2,2}\otimes \mathcal{R}_{2,2}=\left(%
\begin{array}{cccc}
  1 & 1&1&1\\
   1 & -1&1&-1\\
   1 & 1&-1&-1\\
   1 & -1&-1&1\\
\end{array}%
\right),
\end{equation}
such that $\mathcal{R}_{4,4}\in\mathrm{Cor}_{\mathcal{Q}}(4,4)$.
Now, let $X_{k}=\pm 1$ and $Y_{k}=\pm 1$ for all indices
$k=1,2,3,4$. Then we get the following inequality
\begin{eqnarray}\label{chshb}\nonumber
&&\mathrm{X}_{1}(\mathrm{Y}_{1}+\mathrm{Y}_{2}+\mathrm{Y}_{3}+\mathrm{Y}_{4})+
\mathrm{X}_{2}(\mathrm{Y}_{1}-\mathrm{Y}_{2}+\mathrm{Y}_{3}-\mathrm{Y}_{4})\\\nonumber&&+
\mathrm{X}_{3}(\mathrm{Y}_{1}+\mathrm{Y}_{2}-\mathrm{Y}_{3}-\mathrm{Y}_{4})+
\mathrm{X}_{4}(\mathrm{Y}_{1}-\mathrm{Y}_{2}-\mathrm{Y}_{3}+\mathrm{Y}_{4})\leq8.
\end{eqnarray}
 Based on this
inequality, we obtain the generalized CHSH inequality as in the
equation (\ref{chsh1})
\begin{eqnarray}\label{chsh2}
&&|\langle\mathrm{X}_{1},\mathrm{Y}_{1}\rangle_{c}+
\langle\mathrm{X}_{1},\mathrm{Y}_{2}\rangle_{c}
+\langle\mathrm{X}_{1},\mathrm{Y}_{3}\rangle_{c}
+\langle\mathrm{X}_{1},\mathrm{Y}_{4}\rangle_{c}\\\nonumber&&+
\langle\mathrm{X}_{2},\mathrm{Y}_{1}\rangle_{c}-
\langle\mathrm{X}_{2},\mathrm{Y}_{2}\rangle_{c}+
\langle\mathrm{X}_{2},\mathrm{Y}_{3}\rangle_{c}
-\langle\mathrm{X}_{2},\mathrm{Y}_{4}\rangle_{c}\\\nonumber&&+
\langle\mathrm{X}_{3},\mathrm{Y}_{1}\rangle_{c}+
\langle\mathrm{X}_{3},\mathrm{Y}_{2}\rangle_{c}
-\langle\mathrm{X}_{3},\mathrm{Y}_{3}\rangle_{c}
-\langle\mathrm{X}_{3},\mathrm{Y}_{4}\rangle_{c}\\\nonumber&&+
\langle\mathrm{X}_{4},\mathrm{Y}_{1}\rangle_{c}-
\langle\mathrm{X}_{4},\mathrm{Y}_{2}\rangle_{c}
-\langle\mathrm{X}_{4},\mathrm{Y}_{3}\rangle_{c}
+\langle\mathrm{X}_{4},\mathrm{Y}_{4}\rangle_{c} |\leq8.
\end{eqnarray}
Then we can get the following Bell operator with  the same structure
as the combination  which appears on the CHSH inequality as
\begin{eqnarray}
\mathcal{B}_{4,4}&=&\left(%
\begin{array}{cccc}
  \widehat{\mathrm{X}}_{1} & \widehat{\mathrm{X}}_{1}  & \widehat{\mathrm{X}}_{3}
  & \widehat{\mathrm{X}}_{4}  \\
\end{array}%
\right)\mathcal{R}_{4,4}\left(%
\begin{array}{c}
  \widehat{\mathrm{Y}}_{1} \\
  \widehat{\mathrm{Y}}_{2} \\
  \widehat{\mathrm{Y}}_{3} \\
 \widehat{\mathrm{Y}}_{4} \\
\end{array}%
\right)\\\nonumber&=&
\widehat{\mathrm{X}}_{1}(\widehat{\mathrm{Y}}_{1}+\widehat{\mathrm{Y}}_{2}
+\widehat{\mathrm{Y}}_{3}+\widehat{\mathrm{Y}}_{4})+
\widehat{\mathrm{X}}_{2}(\widehat{\mathrm{Y}}_{1}-\widehat{\mathrm{Y}}_{2}
+\widehat{\mathrm{Y}}_{3}-\widehat{\mathrm{Y}}_{4})
\\\nonumber&+&
\widehat{\mathrm{X}}_{3}(\widehat{\mathrm{Y}}_{1}+\widehat{\mathrm{Y}}_{2}
-\widehat{\mathrm{Y}}_{3}-\widehat{\mathrm{Y}}_{4})+
\widehat{\mathrm{X}}_{4}(\widehat{\mathrm{Y}}_{1}-
\widehat{\mathrm{Y}}_{2}-\widehat{\mathrm{Y}}_{3}+\widehat{\mathrm{Y}}_{4}).
\end{eqnarray}
 Now, we will apply the same procedure  as in the case of finding the upper
bound for the violation of the CHSH inequality, by squaring the
Bell operator $\mathcal{B}^{2}_{4,4}$ which is given by
\begin{eqnarray}\label{bell}
&=&\nonumber [\widehat{\mathrm{X}}_{1},\widehat{\mathrm{X}}_{2}]
([\widehat{\mathrm{Y}}_{2},\widehat{\mathrm{Y}}_{3}]
+[\widehat{\mathrm{Y}}_{4},\widehat{\mathrm{Y}}_{3}]
+[\widehat{\mathrm{Y}}_{4},\widehat{\mathrm{Y}}_{1}]
+[\widehat{\mathrm{Y}}_{2},\widehat{\mathrm{Y}}_{1}])
\\\nonumber&&
+[\widehat{\mathrm{X}}_{1},\widehat{\mathrm{X}}_{3}]
([\widehat{\mathrm{Y}}_{4},\widehat{\mathrm{Y}}_{2}]
+[\widehat{\mathrm{Y}}_{4},\widehat{\mathrm{Y}}_{1}]
+[\widehat{\mathrm{Y}}_{3},\widehat{\mathrm{Y}}_{1}]
+[\widehat{\mathrm{Y}}_{3},\widehat{\mathrm{Y}}_{2}])
\\\nonumber&&
+[\widehat{\mathrm{X}}_{1},\widehat{\mathrm{X}}_{4}]
([\widehat{\mathrm{Y}}_{3},\widehat{\mathrm{Y}}_{4}]
+[\widehat{\mathrm{Y}}_{2},\widehat{\mathrm{Y}}_{1}]
+[\widehat{\mathrm{Y}}_{3},\widehat{\mathrm{Y}}_{1}]
+[\widehat{\mathrm{Y}}_{2},\widehat{\mathrm{Y}}_{4}])
\\\nonumber&&
+[\widehat{\mathrm{X}}_{2},\widehat{\mathrm{X}}_{3}](
[\widehat{\mathrm{Y}}_{2},\widehat{\mathrm{Y}}_{4}]+
[\widehat{\mathrm{Y}}_{4},\widehat{\mathrm{Y}}_{3}]+
[\widehat{\mathrm{Y}}_{3},\widehat{\mathrm{Y}}_{1}]+
[\widehat{\mathrm{Y}}_{1},\widehat{\mathrm{Y}}_{2}])
\\\nonumber&&
+[\widehat{\mathrm{X}}_{2},\widehat{\mathrm{X}}_{4}](
[\widehat{\mathrm{Y}}_{2},\widehat{\mathrm{Y}}_{3}]+
[\widehat{\mathrm{Y}}_{3},\widehat{\mathrm{Y}}_{1}]+
[\widehat{\mathrm{Y}}_{1},\widehat{\mathrm{Y}}_{4}]+
[\widehat{\mathrm{Y}}_{4},\widehat{\mathrm{Y}}_{2}])
\\\nonumber&&
+[\widehat{\mathrm{X}}_{3},\widehat{\mathrm{X}}_{4}](
[\widehat{\mathrm{Y}}_{1},\widehat{\mathrm{Y}}_{4}]+
[\widehat{\mathrm{Y}}_{2},\widehat{\mathrm{Y}}_{1}]+
[\widehat{\mathrm{Y}}_{4},\widehat{\mathrm{Y}}_{3}]+
[\widehat{\mathrm{Y}}_{3},\widehat{\mathrm{Y}}_{2}])
\\\nonumber&&
+\{\widehat{\mathrm{X}}_{1},\widehat{\mathrm{X}}_{2}\}(
\{\widehat{\mathrm{Y}}_{1},\widehat{\mathrm{Y}}_{3}\}-
\{\widehat{\mathrm{Y}}_{2},\widehat{\mathrm{Y}}_{4}\})\\\nonumber&&
+\{\widehat{\mathrm{X}}_{1},\widehat{\mathrm{X}}_{3}\}(
\{\widehat{\mathrm{Y}}_{1},\widehat{\mathrm{Y}}_{2}\}-
\{\widehat{\mathrm{Y}}_{3},\widehat{\mathrm{Y}}_{4}\})
\\\nonumber&&
+\{\widehat{\mathrm{X}}_{1},\widehat{\mathrm{X}}_{4}\}(
\{\widehat{\mathrm{Y}}_{1},\widehat{\mathrm{Y}}_{4}\}-
\{\widehat{\mathrm{Y}}_{2},\widehat{\mathrm{Y}}_{3}\})\\\nonumber&&
+\{\widehat{\mathrm{X}}_{2},\widehat{\mathrm{X}}_{3}\}(
\{\widehat{\mathrm{Y}}_{2},\widehat{\mathrm{Y}}_{3}\}-
\{\widehat{\mathrm{Y}}_{1},\widehat{\mathrm{Y}}_{4}\})
\\\nonumber&&
+\{\widehat{\mathrm{X}}_{2},\widehat{\mathrm{X}}_{4}\}(
\{\widehat{\mathrm{Y}}_{3},\widehat{\mathrm{Y}}_{4}\}-
\{\widehat{\mathrm{Y}}_{1},\widehat{\mathrm{Y}}_{2}\})\\&&
+\{\widehat{\mathrm{X}}_{3},\widehat{\mathrm{X}}_{4}\}(
\{\widehat{\mathrm{Y}}_{2},\widehat{\mathrm{Y}}_{4}\}-
\{\widehat{\mathrm{Y}}_{1},\widehat{\mathrm{Y}}_{3}\})+16\mathrm{I}.
\end{eqnarray}
In similarity with the CHSH inequality we can chose
$[\widehat{\mathrm{X}}_{k},\widehat{\mathrm{X}}_{l}]=
[\widehat{\mathrm{Y}}_{k},\widehat{\mathrm{Y}}_{l}]=0$ for all $k$
and $l$, that is, these are commuting observables on both sites. The
result is the following inequality
\begin{eqnarray}
\mathcal{B}^{2}_{4,4}&=& 16\mathrm{I}
+\{\widehat{\mathrm{X}}_{1},\widehat{\mathrm{X}}_{2}\}(
\{\widehat{\mathrm{Y}}_{1},\widehat{\mathrm{Y}}_{3}\}-
\{\widehat{\mathrm{Y}}_{2},\widehat{\mathrm{Y}}_{4}\})\\\nonumber&&
+\{\widehat{\mathrm{X}}_{1},\widehat{\mathrm{X}}_{3}\}(
\{\widehat{\mathrm{Y}}_{1},\widehat{\mathrm{Y}}_{2}\}-
\{\widehat{\mathrm{Y}}_{3},\widehat{\mathrm{Y}}_{4}\})
\\\nonumber &&
+\{\widehat{\mathrm{X}}_{1},\widehat{\mathrm{X}}_{4}\}(
\{\widehat{\mathrm{Y}}_{1},\widehat{\mathrm{Y}}_{4}\}-
\{\widehat{\mathrm{Y}}_{2},\widehat{\mathrm{Y}}_{3}\})\\\nonumber&&
+\{\widehat{\mathrm{X}}_{2},\widehat{\mathrm{X}}_{3}\}(
\{\widehat{\mathrm{Y}}_{2},\widehat{\mathrm{Y}}_{3}\}-
\{\widehat{\mathrm{Y}}_{1},\widehat{\mathrm{Y}}_{4}\})
\\\nonumber &&
+\{\widehat{\mathrm{X}}_{2},\widehat{\mathrm{X}}_{4}\}(
\{\widehat{\mathrm{Y}}_{3},\widehat{\mathrm{Y}}_{4}\}-
\{\widehat{\mathrm{Y}}_{1},\widehat{\mathrm{Y}}_{2}\})\\\nonumber&&
+\{\widehat{\mathrm{X}}_{3},\widehat{\mathrm{X}}_{4}\}(
\{\widehat{\mathrm{Y}}_{2},\widehat{\mathrm{Y}}_{4}\}-
\{\widehat{\mathrm{Y}}_{1},\widehat{\mathrm{Y}}_{3}\}).
\end{eqnarray}
An estimation of this inequality gives
\begin{eqnarray}
&&\mathrm{Tr}(\rho \mathcal{B}_{4,4})\leq8,
\end{eqnarray}
where the observables satisfies $\|\widehat{\mathrm{X}}_{k}\|\leq1$
and $\|\widehat{\mathrm{Y}}_{l}\|\leq1$ for all $k,l=1,2,3,4$.
Moreover, we have supposed that the anticommutators does not vanish
for these observables. Note that
$\widehat{\mathrm{X}}_{k}\widehat{\mathrm{X}}_{l}=\pm\widehat{\mathrm{X}}_{l}\widehat{\mathrm{X}}_{k}$
and
$\widehat{\mathrm{Y}}_{k}\widehat{\mathrm{Y}}_{l}=\pm\widehat{\mathrm{Y}}_{l}\widehat{\mathrm{Y}}_{k}$
 implies $\widehat{\mathrm{Y}}_{k}\widehat{\mathrm{Y}}_{l}=\widehat{\mathrm{X}}_{k}\widehat{\mathrm{X}}_{l}=0$.
If we keep this in mind, then we can get an upper bound of the
maximal quantum violation for the equation (\ref{chsh2}). If we
estimate the inequality without letting any of the observables
commute on both sites, then we get
\begin{eqnarray}
&&\mathrm{Tr}(\rho \mathcal{B}_{4,4})\leq\sqrt{160}=4 \sqrt{10}.
\end{eqnarray}
 Now, we would like to compare this result with Tsirelson upper
 bound for CHSH inequality, where
 $\frac{1}{2}\mathrm{Tr}(\rho
 \mathcal{B}_{2,2})\leq\sqrt{2}\simeq1.41$.
For the generalized CHSH inequality with four observables on both
sites we get
\begin{equation}\label{Cirel}
\frac{1}{8}\mathrm{Tr}(\rho
\mathcal{B}_{4,4})\leq\sqrt{\frac{5}{2}}\simeq1.58,
\end{equation}
where we have used $
\|\{\widehat{\mathrm{Z}}_{k},\widehat{\mathrm{Z}}_{l}\}\|\leq2 ~~
\|[\widehat{\mathrm{Z}}_{k},\widehat{\mathrm{Z}}_{l}]\|\leq2 $ for
$\widehat{\mathrm{Z}}=\widehat{\mathrm{X}},\widehat{\mathrm{Y}}$ and
for all $k,l=1,2,3,4$. This estimation differ from Tsirelson upper
bound for the CHSH inequality because of existence of commutators
and anti-commutators in the square of the Bell operator
(\ref{bell}). However, this is what we expect to get from Tsirelson
idea that quantum correlation should approach the Grothendieck's
constant as the number of observables increase on both sites of a
correlation experiment. Moreover, it is very difficult to show that
these upper bound is tight, that is, the equality is approached for
some quantum state, this needs further investigations.

We can also generalize this result in a straightforward manner
into a generalized CHSH inequality with  $n=m=2^{d}$ dichotomic
observables per site. In this case, we will consider an inequality
that provides a hyperplane separating the polyhedron
$\mathrm{Cor}_{\mathcal{C}}(2^{d},2^{d})$ from the quantum
realizable matrix
\begin{equation}
\mathcal{R}_{2^{d},2^{d}}=\overbrace{\mathcal{R}_{2,2}\otimes\cdots\otimes
\mathcal{R}_{2,2}}^{d},
\end{equation}
such that
$\mathcal{R}_{2^{d},2^{d}}\in\mathrm{Cor}_{\mathcal{Q}}(2^{d},2^{d})$.
Based on this idea, we can get the following Bell operator
\begin{eqnarray}\label{bell2}
\mathcal{B}_{2^{d},2^{d}}&=&\left(%
\begin{array}{cccc}
  \widehat{\mathrm{X}}_{1} & \widehat{\mathrm{X}}_{1}  & \cdots & \widehat{\mathrm{X}}_{2^{d}}  \\
\end{array}%
\right)\mathcal{R}_{2^{d},2^{d}}\left(%
\begin{array}{c}
  \widehat{\mathrm{Y}}_{1} \\
  \widehat{\mathrm{Y}}_{2} \\
  \vdots \\
 \widehat{\mathrm{Y}}_{2^{d}} \\
\end{array}%
\right).
\end{eqnarray}
 Now, we will apply the same procedure as in the case of four observables per site
 by squaring the
Bell operator $\mathcal{B}_{2^{d},2^{d}}$. Then we can write
$\mathcal{B}^{2}_{2^{d},2^{d}}$ in terms of commutator and
anticommutator. However, note that this estimation is only valid for
$d\geq2$ since for $d=1$ we do not have any anticommutator in our
expression for the Bell operator. Next, we  chose
$[\widehat{\mathrm{X}}_{k},\widehat{\mathrm{X}}_{l}]=
[\widehat{\mathrm{Y}}_{k},\widehat{\mathrm{Y}}_{l}]=0$ for all $k$
and $l$. An estimation of this inequality gives
\begin{eqnarray}
\mathrm{Tr}(\rho \mathcal{B}_{2^{d},2^{d}})&\leq&(4
\frac{2^{d}(2^{d}-1)}{2}2^{d-1}+2^{2d})^{\frac{1}{2}}=2^{\frac{3}{2}d},
\end{eqnarray}
where the first term is a contribution from the anticommutators and
second term from the identity operators, which are the squares of
the observables, that is
$\widehat{\mathrm{X}}^{2}_{k}=\widehat{\mathrm{Y}}^{2}_{l}=\mathrm{I}$
for all $k,l=1,2,\ldots,2^{d}$. Now, we can get an upper bound on
the generalized CHSH inequality (\ref{bell2}) if we estimate the
inequality without letting any of the observables commute on both
sites, that is
\begin{eqnarray}\nonumber
\mathrm{Tr}(\rho \mathcal{B}_{2^{d},2^{d}})&\leq&(4
\frac{2^{d}(2^{d}-1)}{2}2^{d}+4
\frac{2^{d}(2^{d}-1)}{2}2^{d-1}+2^{2d})^{\frac{1}{2}}\\&=&2^{d}(3\cdot
2^{d}-2)^{\frac{1}{2}},
\end{eqnarray}
 where the first term is a contribution from the commutators.
 Thus in the general case with $2^{d}$ observables  per site we
 get
\begin{equation}
\frac{1}{2^{\frac{3}{2}d}}\mathrm{Tr}(\rho
\mathcal{B}_{2^{d},2^{d}})\leq 2^{-\frac{d}{2}}(3\cdot
2^{d}-2)^{\frac{1}{2}}.
\end{equation}
Let us analysis this inequality. For CHSH inequality with two
observables per site this estimation does not work since there is no
contribution from anticommutator in this inequality. In the case of
four observables per site, we get the same result as in equation
(\ref{Cirel}) that is  $\frac{1}{2^{3}}\mathrm{Tr}(\rho
\mathcal{B}_{2^{2},2^{2}})\leq\sqrt{\frac{5}{2}}$. And finally, for
a very large number of observables per site, that is whenever
$d\rightarrow\infty$, we have
$\lim_{d\rightarrow\infty}\frac{1}{2^{\frac{3d}{2}}}\mathrm{Tr}(\rho
\mathcal{B}_{2^{d},2^{d}})\leq\sqrt{3}\approx1.73$. This is less
than upper bound for the Grothendieck's  constant ($\approx 1.782$).
However, it almost coincides with the approximate value of
Grothendieck's  constant. Moreover, it can be seen that in our
inequality the maximal quantum violation increases with the number
of observables per site and approaches the maximum value $\sqrt{3}$.

\section{Conclusion}
In this paper, we have constructed an especial type of the CHSH
inequality with four observables per site of a correlation
experiment and we have shown that for arbitrary state the quantum
violation is higher than the Tsirelson bound for CHSH inequality.
Moreover, we have estimated the maximal quantum violation for very
large but equal number  observables  on both sites of a correlation
experiment.  The estimation shows that in this case the maximal
quantum violation converges to $\sqrt{3}\approx1.73$, which
coincides with Grothendieck's constant. This result also can be seen
as an indirect estimation of Grothendieck's constant. However, this
estimation needs further investigation. The approximative value of
this constant was pointed out by Tsirelson \cite{Tsirelson}. In this
paper, he also has discussed the difficultly to find a quantum state
that gives the maximal quantum violation for a given CHSH type
inequality. We should mention that the CHSH inequality does not
include any anticommutator but a generalized CHSH inequality does
include both commutators and anticommutators. In our estimation, we
have assumed that the values of these commutators and
anticommutators do coexist simultaneously and contribute to an
estimation of the maximal quantum violation. We also should mention
that exact value of the Grothendieck is not known yet and our
results could be interesting for the research on this subject.

\begin{acknowledgments}
 The author acknowledge useful discussions with Gunnar Bj\"{o}rk
and Marek Zukowski. The author also would like to thank Jan
Bogdanski. This work was supported by the Wenner-Gren Foundations
and Japan Society for the Promotion of Science (JSPS).
\end{acknowledgments}

\end{document}